\documentstyle[aps,preprint]{revtex}
\begin{document}
\preprint{IMSc-95/34;~hep-th/9603007}
\title{Planckian Scattering from Kerr Black Holes: Eikonal and Beyond}
\author{Saurya Das and R. Parthasarathy
\footnote{E-Mail:~saurya,sarathy@imsc.ernet.in}}
\address{The Institute of Mathematical Sciences, \\ CIT Campus,
Madras - 600 113,  India.}
\maketitle
\begin{abstract}
Planckian scattering of particles with angular momenta is studied
by describing them as sources of Kerr metric. In the shock wave
formalism, it is found that the angular momenta do not contribute
to the scattering amplitude in the eikonal limit. This is
confirmed by using the wave equation of the test particle in the
Kerr background.

\end{abstract}
\newpage

\section{Introduction}

Planckian scattering of point particles in Minkowski space is interesting
because it can be used to probe certain quantum effects of gravity. The
gravitational coupling constant $G$ being of dimension
$(length)^2$, two dimensionless couplings can be constructed out
of it for a system of interacting particles, namely $Gs$ and $Gt$,
where $s$ and $t$ are respectively the squares of the
centre-of-mass energy and the momentum transfer.
Quantum fluctuations of gravitational quantities are expected when
either or both of them are of the order of one (which means that
$\sqrt{s}$ or $\sqrt{t}$ or both approach the Planck energy scale).
In general, the complete analysis of this situation would require a
full theory of quantum gravity. However, since the latter is not
satisfactorily developed so far, one seeks other avenues to
achieve at least partial information of the above. One
interesting kinematical regime is $Gs \sim 1$ and $Gt$ is small
and kept fixed, which is the so-called 
{\it eikonal} limit. Physically, this amounts to vanishingly small
momentum transfers and almost forward scattering. But the merit of
this restriction lies in that, the scattering amplitudes of
point particle scattering 
processes can be calculated {\it exactly}, albeit
semi-classically. Attempts were made in this direction by several
authors \cite{thf,acv,ver,vish}. Basically, one considers the
scattering of a test particle in the `shock-wave' background
provided by the other particle. The interaction is instantaneous
and hence easily tractable.
The formalism has been generalized to
include particles carrying electric and magnetic charges
\cite{dm1,dm2,dm3,dm4}. While the former introduces minor correction terms
to the scattering amplitude for neutral particles 
at the Planck scale, the latter entails contributions comparable
to gravitational ones. One can also try to deviate from general
relativity by introducing metrics which are solutions of say
the low energy equations of motion of string theory \cite{dm5}.

In order to understand the various possibilities that exist in
the above stated kinematical domain, it is natural to examine the
effects of other properties of the particles undergoing
scattering. For example, if they carry angular momentum, then it
would be interesting to see how it modifies the scattering
amplitude of neutral particles. 
In general, we know that angular momenta affects the metric aound
a spherically symmetric matter distribution in a non-trivial
manner \cite{kerr}. The resultant space-time is described by the 
Kerr metric, which exhibits only
axial symmetry. However, in many cases, it is seen that
gravitational effects largely dominate at the Planck scale. 
In this paper, we analyse the situation where 
one of the particles carry a constant orbital angular 
momentum. Inclusion of spin should not be
difficult. The particles being effectively  massless, 
it is reasonable to choose
the direction of Lorentz boost to be parallel (or anti-parallel) 
to the angular
momentum vector. With these specifications, the technique of
shock wave 
scattering can be readily employed. In the next section, we
impart a Lorentz boost on the rotating Kerr metric after a proper
linearization procedure and take the luminal limit. The resulting
gravitational shock wave is seen to deviate from the Schwarzschild
shock wave at a singular point. This gives rise to an interesting
quantization condition involving the angular momentum and the
masses involved. The scattering amplitude for the two particle
process is calculated which reproduces the earlier (non-rotating)
results. In section \ref{external}, we look at the
same scenario from a different perspective. Instead of looking at
the effect of the
speeding rotating black hole, we study the wave equation of a
light test particle in the external field of a rotating matter,
which itself does not undergo any translation. In the eikonal
limit, the scattering amplitude obtained from the boosting
approach is recovered. Moreover, quite remarkably, angular momentum
contributions fail to show up even in the leading order corrections to
the eikonal. This strengthens the previous results and confirms
the supremacy of gravitation at very high energies. The results
are summarized in the concluding section. 

\section{Kerr Shock Wave}

We start with the Kerr metric in Boyer-Lindquist coordinates 
\cite{chandra}:
\begin{eqnarray}
ds^2~&=&~-\left(1-\frac{2GMr}{\Sigma}\right)dt^2 +
\frac{\Sigma}{\Delta}dr^2 + \Sigma d\theta^2  \nonumber \\
&+& \left( r^2 + a^2 +
\frac{2GMra^2\sin^2\theta}{\Sigma}\right) \sin^2\theta d\phi^2 -
\frac{4GMra\sin^2\theta}{\Sigma} dt d\phi~,  
\end{eqnarray}
where
$$\Delta~=~r^2 - 2GMr +a^2~,$$
$$\Sigma~=~r^2 + a^2 \cos^2\theta~,$$
and $a$ being the angular momentum for unit mass.
We model one of the scattering particles of mass $M$
as a source of the above
metric. 
The total angular momentum vector $\vec J$ is along the
$z$-axis. Since we are interested in scattering at large
impact parameters, it would suffice to use the form of the above
metric for large $r$. Expanding the metric in terms of $r$ and
retaining the leading order terms, one gets:
\begin{eqnarray}
ds^2~=~&-&\left(1 - \frac{2GM}{r} \right)~dt^2 + \left(1 +
\frac{2GM}{r} \right)~\left( dx^2 + dy^2 + dz^2 \right) 
 \nonumber \\ 
&+& \frac{4GJ}{r^3} \left( ydx -x dy
\right)~dt~.
\label{metr}
\end{eqnarray}
Here we have substituted $J=Ma$. Note that setting $J=0$ in the last term
reproduces the static Schwarzschild metric. It is interesting to
note that the same metric is obtained by solving the Einstein
equations at large distances 
with a slowly rotating body as the source of a conserved energy-momentum
tensor \cite{papa}. 
Now, since we consider the
centre-of-mass energy of the system to be very high, the relative
velocity between the particles is necessarily very large. For
simplicity, a particular Lorentz frame is chosen, in which the
rotating particle moves at almost the speed of light, while the
other is nearly stationary. From what follows, it will be seen
that this particular choice greatly simplifies the calculations.
To obtain the space-time of the rotating particle as seen from a
stationary observer, we apply the following Lorentz transformation on
the metric (\ref{metr}) along the $z$ - axis 
(i.e. along $\vec J$) \footnote{ one can also consider a boost
perpendicular to $\vec J$, but it is quite unnatural for
luminal velocities.}:
\begin{eqnarray}
x'^0~=~\gamma \left( x^0 + \beta x^1 \right)~, \nonumber \\
x'^1~=~\gamma \left( x^1 + \beta x^0 \right)~, 
\label{lorent}
\end{eqnarray}
and take the limit $\beta \equiv v/c \rightarrow 1$ (or $\gamma
\equiv \left(1 - \beta^2 \right)^{-1/2} \rightarrow \infty$). It is
convenient to work in the lightcone coordinates $x^{\pm} = t \pm
z$. The new metric components will be related to the
original ones by the usual transformation relation:
$$g'_{\mu \nu}~=~\frac{\partial x^{\lambda}}{\partial
x^{\mu}}~\frac{\partial x^\sigma}{\partial x^\nu}~g_{\lambda
\sigma}~.$$
Then to evaluate the singular limit $\beta \rightarrow 1$, we
follow the procedure adopted in \cite{jac}, namely we take the
Fourier transform of the limiting quantity with respect to $x^-$,
then implement the said limit and finally perform an inverse
Fourier transform. Simultaneously, the mass of the particle is
parametrized as $M = p_0/\gamma$, where $p_0$ is the energy of the
luminal particle. 
The calculations are fairly straightforward and
the only surviving components are (dropping the primes):
$$g_{+-} = g_{-+} = -1/2~,$$
$$g_{--} = -8 Gp_0 \ln x_{\perp}~,$$
$$and~~g_{- \phi} = g_{\phi -} = 2 G J~\delta(x^-)~.$$
Here, $x_{\perp} = \sqrt{x^2 + y^2}$ and $\tan \phi =y/x$.
In deriving the above, we have used the following limits
\cite{jac}:
\begin{eqnarray}
&\lim&_{\beta \rightarrow 1}~\frac{1}{R_{\beta}}~=~-2\delta(x^-)~\ln
x_{\perp}~, \nonumber \\
&\lim&_{\beta \rightarrow 1} \frac{1 -
\beta^2}{R_{\beta}^3}~=~\frac{2}{x_{\perp}^2}~\delta(x^-)~,
\nonumber
\end{eqnarray}
where $R_\beta \equiv \sqrt{(x^-)^2 + (1 - \beta^2)^2~x_{\perp}^2}$.
The resultant space-time is described by the invariant line element,
\begin{equation}
ds^2~=~-dx^- \left[~dx^+ + 8Gp_0 \ln x_{\perp}~\delta(x^-)~dx^-
\right] + dx_{\perp}^2 + 4G J~\delta(x^-)
d\phi~dx^-~.
\end{equation}
It is clear that this represents a flat Minkowski space-time
everywhere except on the null plane $x^-=0$, representing the
transverse plane travelling at the speed of light and at the
origin of which the particle resides. Consequently, all curvatures
are localized on this so called `shock-plane'. 
Once again, setting $J=0$ reproduces the shock wave due to a
neutral particle, obtained by boosting the Schwarzschild metric
\cite{aich,thdr}. The above geometry has a simple interpretation
which is borne out by writing it in the form:
\begin{equation}
ds^2~=~-dx^- \left[~dx^+ + 8Gp_0 \ln x_{\perp}~\delta(x^-)~dx^-
- 4G J~\delta(x^-)
d\phi~\right] + dx_{\perp}^2~.
\label{boo}
\end{equation}
Defining the new set of coordinates
\begin{eqnarray}
\tilde x^+~&=&~x^+ + 8G p_0~\theta(x^-)~\ln x_{\perp} + 4GJ \phi~\delta(x^-)~,
\nonumber \\
\tilde x^-~&=&~x^-~, \label{trans} \\
\tilde x_{\perp}~&=&~x_{\perp}~,  \nonumber
\end{eqnarray}
Eq. (\ref{boo}) can be written as,
\begin{equation}
ds^2~=~-d\tilde x^-d\tilde x^+ + d\tilde x_{\perp}^2~.
\end{equation}
Although this represents a flat space-time, the discontinuity in
the transition functions imply that there are
actually two Minkowski spaces glued along $x^-=0$ and the
coordinate $x_{\perp}$ suffers a finite discontinuity every time
there is a crossover from one to the other. Several comments are
in order here. First, there is a non-trivial coordinate shift
along $x^+$ only, the other coordinates being manifestly
continuous. Similarly, if the source particle was
taken to be right-moving and if one started with a left-moving
particle, the the other light cone
coordinate would have suffered an identical discontinuity.
Further, this signifies that for another
fast particle in the above background, the coordinate $x^-$ serves
as a bona fide affine parameter. This has been verified by solving
for the classical null geodesics in presence of the shock-wave
geometry \cite{thdr}. Second, for any test particle, the
sustainable shift in the $x^+$ coordinate when it is hit by the
shock wave is given by the Schwarzschild piece $(8Gp_0 \ln
x_{\perp})$. The second, angular momentum dependent piece has 
support only for an infinitesimal time 
around $x^-=0$ and vanishes everywhere
else. Thus its effect is not retained by the test particle at
later times. We shall see, however, that it has other interesting
consequences. It may be noted that the gravitational shock waves
in the context of Kerr geometry has been considered earlier in
\cite{ls}. However, there the {\it total} angular momentum $J$ was
also scaled as $ J \rightarrow J/\gamma$. This means that in the
singular limit $\gamma \rightarrow \infty$, $J$ simply decays to
zero. On the other hand, we have chosen to examine the implications 
of keeping $J$ fixed at some finite value. The motivation is of
course 
that in reality, there can exist particles travelling at the
velocity of light and carrying finite orbital or
spin angular moment and we would like to determine its role in Planckian
scattering. 

Having obtained the shock wave geometry of a 
ultra-relativistic rotating particle, we now proceed to compute
its effect on the slow test particle. Before the shock wave hits
this particle, its wave function can be taken to be a plane wave
carrying momentum $p'$.
(without loss of generality, we take this particle to be spinless),
\begin{eqnarray}
\psi_<~&=&~\exp [ip'x] \nonumber \\
&=& \exp {i[~\vec p~' \cdot \vec x_{\perp} - \frac{1}{2}p'^- x^+ -
\frac{1}{2} p'^+ x^-]} \label{wave}~.
\end{eqnarray}
The impact of the Kerr shock wave introduces an additional phase
in the above, which can be obtained by using in the
shift (\ref{trans}) in ({\ref{wave}). The resultant
wave function is, 
\begin{equation}
\psi_>~=~
 \exp {i[~\vec p~' \cdot \vec x_{\perp} - \frac{1}{2}p'^- \{~ x^+ +
8 G p_0
\ln x_{\perp} + 4 GJ \phi~\delta(x^-)~\} 
- \frac{1}{2} p'^+ x^- ]} \label{wave1}~.
\end{equation}
Concentrating on the shock plane, we see that the new phase factor
is dependent on the azimuthal angle $\phi$.
Thus, the phase picked up by
the test particle wave function on traversing the shock plane,
identified by $\delta (x^-)$, is:
$$ \Phi~=~\frac{Gs J \phi}{p_0}~\delta(x^-)~,$$
on using 
the relation $2p_+p_0 = s$. 
Introducing the test particle mass $m'$ through the
relation $s=2m'p_0$, the phase factor becomes:
$$\Phi~=~4\pi G J m' \delta(x^-)~.$$
for a closed loop around the $z$-axis on the shock plane. 
The condition for single valuedness of the wave function implies
the following quantization condition:
\begin{equation}
4 \pi G J m'~=~n L~,
\end{equation}
where $n$ is an arbitrary integer and $L$ is a length scale
introduced for dimensional consistency.
The physical
significance of this is not very clear as it implies
a quantum condition on the specific combination of the mass as
well as the angular momentum appearing in the equation above. 

Now we proceed to calculate the scattering amplitude in the
eikonal limit. As stated earlier, the interaction is totally
encoded in the logarithmic part of the phase shift $-Gs \ln
x_{\perp}$. Expanding $\psi_>$ in terms of plane waves and
performing an inverse Fourier transform a l\'a ref. \cite{thf} results
in the following expression for the amplitude :
\begin{equation}
f(s,t)~=~\frac{Gs}{t}~\frac{\Gamma \left(1-iGs\right)}{\Gamma
\left(1+iGs\right)} \left( \frac{1}{-t}\right)^{-iGs}~,
\label{scat}
\end{equation}
upto standard kinematical factors. Observe that it resembles the
Rutherford scattering formula with the electromagnetic coupling
constant $\alpha$ replaced by $-Gs$. Thus we arrive at the
important conclusion, that in the eikonal kinematical
regime, the scattering
amplitude is {\it independent} of the angular momenta of the
interacting particles. This was not obvious a priori and is specific of
the limits imposed on the energy-momenta of the particles involved. 
We will confirm this result by a different
approach in the next section where we will also probe its status
away from the eikonal limit. 

\section{Scattering in Kerr background}
\label{external}
We have seen that the rotating particle in the 
infinitely Lorentz boosted frame gave rise to a gravitational
shock wave which produced an instantaneous interaction with the
otherwise free particle. The same situation can also be
visualized in a reciprocal manner, whereby the test particle
travels at the velocity of light past the particle carrying
angular momentum and which does not change its position with time.
As before, the relative velocity is parallel
to the angular momentum vector. The wave function $\psi$ of the
spinless test particle will now satisfy the massless Klein-Gordon equation
in the Kerr background:
\begin{equation}
D_\mu D^\mu \psi~=~\frac{1}{\sqrt{-g}}~\partial_\mu \left( \sqrt{-g}~g^{\mu
\lambda} \partial_\nu~\psi\right)~=~0~,
\label{kg}
\end{equation}
where $D_\mu$ signifies the gravity covariant derivative. The
components of the metric tensor being time independent, 
$\psi$ can be separated as :        
$$\psi (\vec r, t)~=~\psi (r)~e^{iEt}~.$$
Next, to evaluate the D 'Alembertian operator in the eikonal limit,
we linearize metric (\ref{metr}) and write it in the form:
$$g_{\mu \nu}~=~\eta_{\mu \nu} + h_{\mu \nu}~,$$
where the first term on the right represents the Minkowski metric,
while the second term measures the deviations from it. It is
assumed that $|h_{\mu \nu}| \ll |\eta_{\mu \nu}|$, since we are
considering weak gravitational fields.  From Eq.(\ref{metr}) we
find that 
the only non-trivial
$h_{\mu \nu}$ s (in polar coordinates) are
$$h_{00}~=~h_{rr}~=~\frac{2GM}{r}~,~~~
h_{0 \phi}~=~\frac{2GJ \sin^2\theta}{r}~.$$
The corresponding non-vanishing contravariant components
are found to be
$$h^{00}~=~h_{00}~,~~~h^{rr}~=~h_{rr},~~~h^{0 \phi}~=~-\frac{2GJ}{r^3}~.$$
In the linearized approximation, the determinant of the metric
tensor can
be written as:
\begin{equation}
\sqrt {-g}~=~e^{\frac{1}{2} Tr \ln (g_{\mu \nu})}~=~|\eta| - Tr \left(
\eta^{-1} h~\right) + {\cal O}(h^2)~,
\end{equation}
where, $\eta$ and $h$ are $4 \times 4$ matrices. Explicit
calculation shows that the above trace vanishes and $\sqrt{-g} =
r^2 \sin^2 \theta$. 
Using the above results, we get from in (\ref{kg}),
\begin{equation}
\left[~r^2  \left( 1 +
\frac{2GM}{r}\right) E^2 + \partial_r \left\{r^2
\left(1 -\frac{2GM}{r}\right) \partial_r \right\}  
- \hat L^2
- \frac{4GJE}{r}~\hat L_z~ \right]
\psi~=~0~,
\label{kg1}
\end{equation}
where
$$\hat L^2~\equiv~-\left[ \frac{1}{\sin\theta}~\partial_\theta \left(
\sin \theta~\partial_\theta \right) +
\frac{1}{\sin^2\theta}~\partial^2_\phi \right]$$
and
$$\hat L_z=-i\partial_\phi$$
are the square of the angular momentum operator and its
$z$-component respectively in coordinate basis. The operators
acting on $\psi$ in (\ref{kg1}) commute with $\hat L^2$ and $\hat
L_z$ and hence 
$\psi$ can be chosen to be their simultaneous
eigenfunction. In particular, we can choose it in the form:
\begin{equation}
\psi (\vec r, t)~=~\frac{f(r)}{r}~e^{iEt}~
Y_{lm}\left(\theta,\phi\right)~.
\label{psi}
\end{equation}
Thus, although the Kerr metric is axisymmetric, the Klein-Gordon
wave function can be entirely separated into radial and angular
components. This was first shown explicitly in
\cite{teu} using the Newman-Penrose formalism. Here, we have given
simple a group-theoretic argument in its favour in the
linearized gravity approximation.
With this, the radial part of Eq.(\ref{kg1}) becomes,
\begin{eqnarray}
&{}&\frac{d^2f}{dr^2} + \frac{2GM}{r} \left( 1 + \frac{2GM}{r} \right)
\frac{df}{dr} \nonumber \\
&-& \left[\frac{l(l+1) - 3 (Gs)^2}{r^2} -
\frac{2GsE}{r} -E^2 + \frac{2GM l (l+1)}{r^3} + \left( 1 +
\frac{2GM}{r} \right) \frac{4GJmE}{r^3} \right]~f~=~0~.
\label{rad}
\end{eqnarray}
Taking the limit $M\rightarrow 0$, we get:
\begin{equation}
\frac{d^2f}{dr^2} 
- \left[\frac{l(l+1) - 3 (Gs)^2}{r^2} -
\frac{2GsE}{r} -E^2 \right]~f - \left[ \frac{2GM l (l+1)}{r^3} + 
 \frac{4GJmE}{r^3} \right]~f~=~0~.
\label{rad1}
\end{equation}
Here we retain the term $2GM l(l+1)/r^3$ because although $M$ is
small, the angular momentum $l = b E$ is very high (where $b$ is the
impact parameter) and the product may be finite. However, 
the eikonal approximation corresponds to the terms in the first
square brackets which gives the following exact phase shift in the $l
\rightarrow \infty$ limit \cite{dm4} :
\begin{equation}
\delta_l^{eik}~=~\arg \Gamma( l + 1  - iGs)~.
\end{equation}
The scattering amplitude can be obtained by using the familiar relation
:
\begin{equation}
f(s,t)~=~\frac{1}{2i \sqrt s} \sum_{l=o}^{\infty} (2l+1) [
e^{2i\delta_l} - 1 ] P_l (\cos\theta)~,
\label{partial}
\end{equation}
which yields the amplitude (\ref{scat}). Thus, once again we 
arrive at the conclusion that angular momentum does not have any
effect on Planckian gravitational scattering in the eikonal limit.

Next, we proceed to calculate the corrections to the above phase
shift due to the ${\cal O} (1/r^3)$ terms in the radial equation.
Once we include these terms, there is no exact solution of the
wave equation. However, the phase shifts that they induce can be
estimated by the following formula for
a short ranged potential $V(r)$ (which falls off faster than
$1/r^2$ as $r \rightarrow \infty$) \cite{baym}: 
\begin{equation} \delta_l \approx 2ME\int_{0}^{\infty} dr~r^2 V(r)
j_l(kr)\frac{R_l(r)}{r}~. \label{born}
\end{equation}
Substituting $V(r) = 2GM l(l+1) /r^3$, we 
obtain the phase shift  \cite{dm4}
\begin{equation}
\delta_l \approx GM^2~,
\end{equation}
which vanishes for massless particles. 
For the potential $U(r) \equiv 4GJmE/r^3$, 
the argument is more subtle since it
contains the azimuthal quantum number $m$. The expression for the
Born amplitude is 
given by \cite{baym}:
\begin{equation}
f_{\vec k}(\Omega)~=~-\int d^3r'~e^{-i\vec k \cdot \vec
r'}~U(r')~\psi_{\vec k}(\vec r')~.
\end{equation}
Note that for almost forward scattering, this amplitude will be
strongly peaked around $\vec k$ parallel to the $z$-axis. We are
thus justified in replacing $\vec k \cdot \vec r'$ by $kz'$.
Substituting  Eq.(\ref{psi}), and using the formula
$$e^{ikz}~=~\sum_{l=0}^{\infty} i^l (2l+1) P_l(\cos \theta)
j_l(kr)$$ 
we get,
\begin{equation}
f_{\vec k}(\Omega)~=~-\sum_{l=0}^{\infty}~(2l+1)~\int dr~r^2~
U(r)~\frac{f(r)}{r}~j_l(kr)
~i^l \int d\theta \sin \theta P_l(\cos\theta)  
\int d\phi ~Y_{lm}(\theta,\phi)~,
\end{equation}
where $j_l$ is the spherical Bessel function. The integration over
$\phi$ produces the Kronecker delta $\delta_{m0}$. Thus, 
$U(r)$ being proportional to $m$, we conclude that it does not
contribute to the scattering amplitude at all ! This result is
quite remarkable in the sense that the conclusions of the previous
section can be extended to the leading order correction term and
thus establishing the former on a
much firmer footing. Thus, in the eikonal kinematical regime, one
is justified in ignoring quantities such as charge \cite{dm2} and
angular momenta, and knowledge about the energy of the particles 
alone suffices in
calculating the exact scattering amplitude. 

\section{Conclusions}

In summary, our results reveal the importance associated
with the gravitational scattering amplitude (\ref{scat}), first
calculated in ref.\cite{thf}. We found that for all practical
purposes, the gravitational
shock wave of a particle does not bear any signature of its 
angular momentum. Therefore, the latter does not affect the interaction
with another light particle. Similar conclusions follow for the
particles carrying charge, where too the form of the shock wave
remains unaltered \cite{dm2}. This was first anticipated in
\cite{thdr}, but the corresponding proofs are far from trivial. Next,
we show that the same result can be derived {\it ab initio} by
considering the scattering of a test particle from a static
Kerr black hole. Moreover, this formalism has been 
generalized to the leading order correction to the eikonal. 
We do not expect any significant
deviations from our results for incident particles with intrinsic
spin.
However, it
would be interesting to investigate the general case when the
scattering particles carry both spin and orbital angular momenta,
and see whether something analogous to spin-orbit coupling takes
place. We hope to report on this in future.

\begin{center}
{\bf ACKNOWLEDGEMENTS}
\end{center}

\noindent
We would like to thank P. Majumdar and K. S. Viswanathan for
useful discussions.

\end{document}